\documentstyle[aps,preprint,epsfig]{revtex}

\begin{document}
\draft
\def\be{\begin{equation}}
\def\ee{\end{equation}} 
\def\bfi{\begin{figure}}
\def\efi{\end{figure}}
\def\bea{\begin{eqnarray}}
\def\eea{\end{eqnarray}}
\title{The segregation of sheared binary fluids in the Bray-Humayun model}
%\title{The segregation process of binary fluids in an external flow}

\author{Antonio Lamura,  Giuseppe Gonnella}

\address{Istituto Nazionale per la  Fisica della Materia, Unit\`a di Bari
{\rm and} Dipartimento di Fisica, Universit\`a di Bari, {\rm and}
Istituto Nazionale di Fisica Nucleare, Sezione di Bari, via Amendola
173, 70126 Bari, Italy.}

\author{Federico Corberi}

\address{Istituto Nazionale per la Fisica della Materia,
Unit\`a di Salerno and Dipartimento di Fisica, Universit\`a di Salerno,
84081 Baronissi (Salerno), Italy}

\maketitle
 
\begin{abstract}
The phase separation process which follows a sudden quench 
inside the coexistence region is considered for a binary
fluid subjected to an applied shear flow.
This issue is studied in the framework of the convection-diffusion equation
based on a Ginzburg-Landau free energy functional in the
approximation scheme introduced by 
Bray and Humayun [{\it Phys.Rev.Lett.} {\bf 68}, 1559, (1992)].  
After an early stage where domains form and shear effects become
effective the system enters a scaling regime where the typical domains
sizes $L_\parallel $,
$L_\perp$ along the flow and perpendicular to it
grow as $t^{5/4}$ and $t^{1/4}$. The structure factor is characterized
by the existence of four peaks, similarly to previous theoretical
and experimental observations, and by exponential tails at large wavevectors.
\end{abstract}
 

\section{Introduction} \label{intro}

The process of phase-separation occurring in binary systems 
quenched below the coexistence line has been
matter of intensive investigations since many years. The 
overall phenomenology is nowadays reasonably well understood.
In an early stage, whose properties depend on the system specific
details, domains of the two phases form; the
relative concentration of the two species well inside 
the domains quickly  saturates to a value very similar to the
equilibrium composition. Then a crossover leads to the late stage
dynamics which is characterized by the growth of ordered 
regions of typical size $L(t)$ 
and by a statistically invariant morphology.
This phenomenology is at the basis of the well known
scaling property according to which a single length $L(t)$
is asymptotically relevant: system configurations at
different times are simply related by a scale transformation
with a scale factor $L(t)$. In particular, for the equal time
density-density correlation function one has
\be
G(\vec r,t)=g\left [\frac{r}{L(t)}\right ], 
\label{sca}
\ee
meaning precisely that $G$ is invariant
as a function of scaled lengths $r/L(t)$.
The growth of $L(t)$ is of the power law type $L(t)\sim t^{1/z}$.
The dynamic exponent $z$ depends on the specific microscopic mechanisms 
responsible for the segregation process; in general different
dynamical regimes can be observed, characterized by different
values of $z$, by changing the experimental time window \cite{bray}.
When the system enters the late stage the memory of the initial state
is lost and, due to the presence of one single growing length,
its properties become universal. Then the dynamical exponents,
the scaling function $g$ and other observables become independent on 
the system specific details and are only determined by a few
relevant parameters. Besides the spatial dimensionality, 
the presence of conservation laws is a relevant feature. 
For binary fluids, to which we restrict our attention,
the concentration of the species is conserved as the phase-separation
proceeds.  
The initial stage of the scaling
regime, before hydrodynamic effects become relevant, is the subject
of this paper. For viscous fluid this time window
can be sufficiently wide to be detected experimentally.
In this condition the phase-separation process is ruled 
by the evaporation, diffusion and subsequent recondensation 
of monomers \cite{ls} and the exponent $z$ is known to
be $3$, independent on the space dimension $d$. 

The spatial properties are encoded into the
scaling function $g(x)$ or, equivalently, into its Fourier transform 
$f(q)$. For large $q$, $f$ has a power-law tail of the form
$q^{-(d+1)}$: This is the famous Porod's law \cite{porod},
long recognized as arising from sharp domains walls.

Another relevant parameter affecting the scaling properties
is the number of components $N$  of the order parameter $\phi $.
For binary fluids $\phi$ is a scalar field. Most of the analytical
techniques developed to study the kinetics, among which the
Bray-Humayun scheme that is considered throughout this article,
are however suited for vectorial systems with large-$N$. 
Let us briefly discuss the main features of vectorial systems that
will be recalled in this paper.     
For systems with
a continuous symmetry, rotations of the order parameter on the
${\cal O}(N)$ symmetric manifold of the local ground state of
the Hamiltonian, instead of evaporation and condensation, 
is the important growth mechanism. This changes $z$ to $4$, regardless
both of $d$ and $N$ (apart from exceptional cases \cite{bray}).
Concerning the large-$q$ behavior of $f$ a generalization
of the Porod's law, $f(q)\sim q^{-(d+N)}$ is obeyed \cite{bpt}
for $N\le d$. This form relies on the existence of
stable localized topological defects. Much less is known about 
the tails of $f$ when these defects are not present, for
$N>d$. Numerical simulations \cite{rao} found a stretched-exponential
decay $f(q)\sim \exp [-q^\nu ]$.

Despite this rather simple phenomenology the problem of deriving
the scaling properties from first principles is still
open. On the analytical side, the only soluble model \cite{cz} for conserved
order parameter is the one with $N=\infty$. When considering
this unphysical limit one hopes that the gross features of 
the kinetic process are retained for $N=\infty$ 
leaving further refinements to $1/N$ perturbation theories.
However, although the global behavior of the $N=\infty$ model
is qualitatively resemblant \cite{alcuni} of real systems, 
the scaling symmetry~(\ref{sca}) is violated already
at a qualitative level. The explicit solution
of the model \cite{cz} shows in fact that instead of the standard 
scaling~(\ref{sca})
one has a {\it multiscaling} symmetry produced by the existence
of two logarithmically different lengths. Actually the ordering
process with $N=\infty$ is more a {\it condensation} in momentum
space \cite{ccz}, than a genuine phase-separation process.
These results shadow the possibility to develop a successful
theory of the segregation process
by systematic $1/N$ expansions around the large-$N$ limit.
 
In this scenario a prominent role is played by a class of approximate
theories known as gaussian auxiliary field theories, originally
introduced by Mazenko \cite{maz}.
The essence of these is a non-linear mapping between the order
parameter $\phi (\vec x,t)$ and an auxiliary field $m(\vec x,t)$.
The actual form of the relationship $m(\phi)$ is
largely arbitrary and, depending on different physically motivated choices, 
different schemes have been developed \cite{zd}. 
The basic idea is that with a proper choice of $m(\phi)$ the equation
obeyed by the auxiliary field can be treated to lowest order 
in a mean-field approximation
while the basic non-linearities of the evolution are still retained
through the non-linear mapping $m(\phi)$. 
This {\it ad hoc} assumption is generally uncontrolled and its use
is justified {\it a posteriori} by the satisfactory 
quality of the predictions.  
Starting from the method of
Mazenko, Bray and Humayun (BH) have derived \cite{bh} a
non-linear closed equation of motion for the correlation 
function $G(\vec r,t)$ within the framework of the $1/N$ 
perturbative expansion. Letting $N\to \infty$ in the BH equation one
recovers the large-$N$ model previously discussed; the
BH approach, however, does not amount to a straightforward $1/N$-expansion
because all orders in $1/N$ are reintroduced
through the gaussian auxiliary field approximation.
On one side, the uncontrolled character is the limit of the BH
scheme, since it is not clear how to improve the approximation, but 
on the other side, represents a powerful tool to overcome the
shortcomings of standard perturbation theory previously discussed.
As shown by BH their equation reinstates standard scaling~(\ref{sca})
as a truly asymptotic symmetry, 
leaving to multiscaling a merely pre-asymptotic
role. An additional feature at variance
with the $N=\infty$ limit is the form of the scaling function $f(q)$,
which, in the BH approximation falls exponentially at large
momenta, as opposed to the quartic exponential decay of the 
$N=\infty $ model.

In this paper we are interested in the diffusive phase separation process
of a binary fluid in the presence of an
applied shear flow, namely an imposed fluid velocity along $x$
with a constant gradient along $y$. The presence of the flow
radically changes the behavior and even the phenomenology
is much less understood. A first trivial effect is the
anisotropic deformation of the growing pattern that appears
greatly elongated along $x$ \cite{onuki,giap,otha,stat}. 
Then, when dealing with
growth laws one has to specify the length of the equilibrated regions
for each direction, $L_x(t), L_y(t), L_z(t)$, and in general 
$L_x(t)>>L_y(t),L_z(t)$. A more subtle point, that may have
important consequences, is the following:
stretching of the domains causes ruptures of the network \cite{otha}
that may render the segregation incomplete.
Actually, the characteristic lengths in the different directions
could keep on growing indefinitely, as without shear, 
or due to domains break-up the system may eventually
enter a stationary state characterized by domains with a
finite thickness. In general there is no agreement on this point
both from the experimental \cite{stat} and theoretical point of view. 
On the theoretical side large-$N$ calculations \cite{brayninf,noiinf}
show the existence of a multiscaling regime with ever growing
lengths in all directions. However numerical simulations \cite{noinum} 
cannot still establish a clear evidence due to finite size and 
discretization effects.
Another intriguing feature which is probably related to the stretching
and break-up of the domains is the observation of an oscillatory
pattern in the observables. Let us consider the excess viscosity 
$\Delta \eta$ as an example. Stretching of domains require work against surface
tension leading to an increase of $\Delta \eta$. However when a domain
breaks up the stress is released since the fragments are more isotropic; this
produces a decrease of $\Delta \eta$. If repeated in time this 
mechanism produces
an oscillatory pattern that is observed in simulations \cite{noinum}
and possibly in experiments \cite{osc}. In the large-$N$ limit it can be shown
that this process has a periodicity in $\log t$. 
Similar patterns are also reported in apparently different contexts
as fractures in heterogeneous solids \cite{27ninflungo}
or in stock market indices \cite{28ninflungo}, suggesting a
possible common underlying interpretation.
However we do not have nowadays analytical tools
to enlighten this feature.   

The spatial properties of the mixture are encoded into the
correlation function $G(\vec r,t)$ or, equivalently, into its
Fourier transform $C(\vec k,t)$, the structure factor.
The latter is directly measured with scattering techniques 
and is more suited for comparison with the experiments.
The form and the evolution of $C$ are themselves modified
by the presence of the flow. When the shear becomes effective,
after an initial isotropic regime where $C$ has the shape of a circular
volcano, the tip of $C$ is deformed into an ellipses.
This is expected as a consequence of the anisotropy, and is
generally observed in experiments. In some cases the 
presence of four pronounced peaks on the edge of $C$ 
can be detected \cite{exp4picchi}. 
The existence of these multiple maxima has been also
observed in numerical studies \cite{noinum} of phase separation
and in different physical systems such as stationary microemulsion 
in shear flow \cite{micro}, suggesting that this is a rather general feature
of sheared fluids.  
A peak in the structure
factor is generally interpreted as the signature of a characteristic
length proportional to the inverse of the wavevector
where the peak is located. In this anisotropic case to each peak
one associates one typical length for each space direction.
Taking into account the symmetry $\vec k\to -\vec k$ 
this suggests that
a couple of characteristic lengths for each direction is
present: It is not completely clear, however, to which extent 
this interpretation can be pushed.

In this partly unclear scenario the development of  
theoretical tools for reliable predictions to be tested in
experiments is important. 
In the case without shear the BH scheme turns out to be
a reference theory capable to describe the main ingredients
of the phenomenology. Therefore in this paper we address a
numerical solution of the BH equation in the presence of
shear flow. We will compare its behavior 
with previous results pertaining to the large-$N$ 
limit or to numerical simulations of the fully non-linear scalar
model. Our results confirm the occurrence of a scaling
regime with growing lengths in every direction. The dynamical
exponents are found to coincide with those of the $N=\infty$
model. The four peak pattern for the structure factor is
exhibited, similarly to what observed both for $N=\infty $ and 
in the scalar model,
the actual form of $C$ being, however, rather different in the
three cases.
Concerning the oscillating behavior of the typical lengths and
of the excess viscosity, on the other hand, the BH model
behaves quite differently, in that these are strongly inhibited 
or even absent, at variance with $N=\infty$ and $N=1$.

This paper is divided in 4 Sections: In Sec.~\ref{model} we introduce the
BH equation and discuss some numerical details about its integration;
in Sec.~\ref{results} the results of the numerical solution are presented
and a comparison with different approaches is proposed.
In Sec.~\ref{conclusions} we draw the conclusions of this work.
  
\section{The model} \label{model}

We consider a system described by an order parameter with $N$ components.
The evolution is described by the convection-diffusion 
equation \cite{bray} that generalizes
the Cahn-Hilliard-Cook equation in the case of an applied
velocity field $\vec v$
\be
\frac{\partial  {\bf \phi} (\vec x,t)}{\partial t}+\vec \nabla \cdot \left [
{\bf \phi} (\vec x,t)\vec v (\vec x,t)\right ]=\Gamma \nabla ^2 
\frac {\delta {\cal F}\left (\{ {\bf \phi} \} \right
)}{\delta {\bf \phi }}+
{\bf \eta }(\vec x,t)
\label{chc}
\ee
where ${\bf \phi}\equiv \{ \phi _\alpha \}$ 
is the vector order parameter that
in the scalar case represents the concentration difference between the two
species of the fluid, $\vec x\equiv \{x,y,z\}$ is the space coordinate,
$\Gamma $ is a transport coefficient
and ${\bf \eta  }$ is a gaussian stochastic field
with expectations
\begin{eqnarray}
\langle {\bf \eta }(\vec x,t)\rangle &=& 0 \nonumber \\
\langle  {\bf \eta }(\vec x,t)  {\bf \eta }(\vec x',t') \rangle &=&
-2T\Gamma \nabla ^2 \delta (\vec x-\vec x')\delta (t-t')
\label{correnoise}
\end{eqnarray}
describing thermal fluctuations. 
The symbol $\langle \dots \rangle$ means an ensemble average.
${\cal F}$ is the equilibrium
Ginzburg-Landau free energy
\be
{\cal F} (\{ {\bf \phi } \})=\int d\vec x \left [ 
\frac{r}{2}{\bf \phi }^2+\frac{g}{4N} \left ( {\bf \phi }^2\right )^2
+\frac{u}{2}\vert \nabla {\bf \phi} \vert ^2
\right ]
\label{freen}
\ee
where $\phi^2 = \sum_{\alpha} \phi_{\alpha} \phi_{\alpha}$.
The parameter $r$ distinguishes between the mixed state with
$r>r_c(T)$ ($r_c(T)<0$) and the phase separated states 
with $r<r_c(T)$.
For a plane shear flow the velocity term is given by
\be
\vec v=\gamma y \vec e_x
\label{shear}
\ee
where $\vec e_x$ is the unit vector in the flow direction $x$ 
and $\gamma $ is the shear rate.
   
In the approximation framework developed by BH \cite{bh}
the following equation of motion can be derived for the equal time 
correlation function $G(\vec r,t)=\langle \phi _\alpha (\vec x,t)
\phi _\alpha (\vec x+\vec r,t) \rangle$
\be
\frac {\partial G(\vec r,t)}{\partial t}
+\gamma y \frac{\partial G(\vec r,t)}{\partial x}=
-2\nabla ^2\left \{ \nabla ^2 G(\vec r,t)-R(t)\left [
G(\vec r,t)+\frac{1}{N}G^3(\vec r,t) \right ]
\right \} 
\label{bh}
\ee
where we have dropped the component indices due to internal symmetry.
In deriving Eq.~(\ref{bh}) from Eq.~(\ref{chc}) we have let 
$\Gamma =1$ and $u=1$ since this simply amounts to a redefinition
of time and space scales and we have neglected the thermal disturbance
$\eta$ because it can be shown \cite{brayren,noinum} 
that the temperature is an irrelevant parameter below
the coexistence line.
The quantity $R(t)$ is a function of time that is asymptotically 
determined by the requirement $\lim _{t\to \infty} G(0,t)=S_{eq}=-r/g$.
Here we use the self-consistent determination of $R(t)$
\be
R(t)=g[G(0,t)-S_{eq}] 
\label{r}
\ee
that is usually adopted \cite{bh,zancast}.
For $N=\infty$, namely dropping the cubic term on the r.h.s., Eq.~(\ref{bh})
becomes the $N=\infty$ model.

We have considered the evolution of Eq.~(\ref{bh}) 
with $N=1$. 
Some comments about the role of $N$ are in order.
Eq.~(\ref{bh}) is derived in an $1/N$ perturbative framework.
Then, in principle, the quality of the approximation is expected 
to be better for large $N$ and, in any case, this whole approach
is meaningful for vectorial systems with continuous symmetry. 
Given that the nature of the symmetry
is relevant for the scaling properties, as discussed in Sec.~\ref{intro}, 
one cannot pretend to push this
scheme down to the physically relevant case with $N=1$ in a completely 
successful way. However, besides numerical simulation that are
quite difficult and not particularly instructive in this case,
most of the understanding of phase separation with shear comes
nowadays from the $N=\infty$ model, which surely suffers of much
more profound shortcomings than the present approach, 
if used to infer the properties of physical systems.
The BH scheme represents
a first step beyond $N=\infty$, despite of course being not conclusive.
It must be recalled, moreover, that at least without shear 
the BH equation substantially reproduces the behavior of the system
with $N=\infty$, including multiscaling, in a preasymptotic time
domain $t<t^*(N)\sim (\Delta N)^{4/d}(\ln N)^3$ after which
the cubic correction in Eq.~(\ref{bh}) becomes effective \cite{bh}.
The choice $N=1$, besides being appropriate for physical systems, 
is also the most efficient in order to amplify as much as possible the role of 
finite-$N$ corrections, given the limited timescale numerically accessible.

Eq.~(\ref{bh}) have been solved on a two dimensional lattice of size $L=4096$, 
the lattice constant being $\Delta x=1$, via a finite difference first order
Euler scheme with $\Delta t=0.01$. Periodic boundary conditions have been
adopted.
The initial condition $G(\vec r,0)=\Delta \delta (\vec r)$, 
$\Delta $ being a constant, corresponds
to a quench from an equilibrium state at infinite temperature.
The parameter of the free energy we use are $-r=g=1$; different choices 
correspond to a redefinition of the order parameter amplitude.
We will present data relative to the case $\gamma =0.01$ and $\Delta =0.01$;
we have tested that other choices produce similar results.

The structure factor $C(\vec k,t)=\langle \hat \phi (\vec k,t) 
\hat \phi (-\vec k,t)\rangle $ obeys the equation
\be
\frac{\partial C(\vec k,t)}{\partial t}
-\gamma k_x \frac{\partial C(\vec k,t)}{\partial k_y} 
=-2k^2 \left [ k^2 +R(t)\right ]C(\vec k,t)-2\frac{k^2}{N}R(t)D(\vec k,t)
\label{ck}
\ee
where $D(\vec k,t)$ is the Fourier transform of $G^3(\vec r,t)$.
Due to the structure of this Equation, however, it is more efficient to
compute $G$ from Eq.~(\ref{bh}) and to obtain $C$ by
Fourier transform.
From the knowledge of $C$ one can extract the characteristic lengths
along $x$ and $y$ as
\begin{eqnarray}
L_x(t) &=& 
\left (
\frac{\int d\vec k \vert k_x \vert C(\vec k,t)}
{\int d\vec k C(\vec k,t)}  
\right )^{-1} \nonumber \\
L_y(t) &=& 
\left (
\frac{\int d\vec k \vert k_y \vert C(\vec k,t)}
{\int d\vec k C(\vec k,t)}  
\right )^{-1} 
\label{radii}
\end{eqnarray}
and rheological indicators such as the shear stress
\be
\sigma_{xy} (t)=\int \frac{d\vec k}{(2\pi)^d} k_x k_y C(\vec k,t)
\label{stress}
\ee
For a steady flow the excess viscosity is defined in terms of the
shear stress as $\Delta \eta (t)=-\gamma ^{-1} \sigma _{xy}(t)$.

\section{results} \label{results}

\begin{figure}
\centerline{\psfig{figure=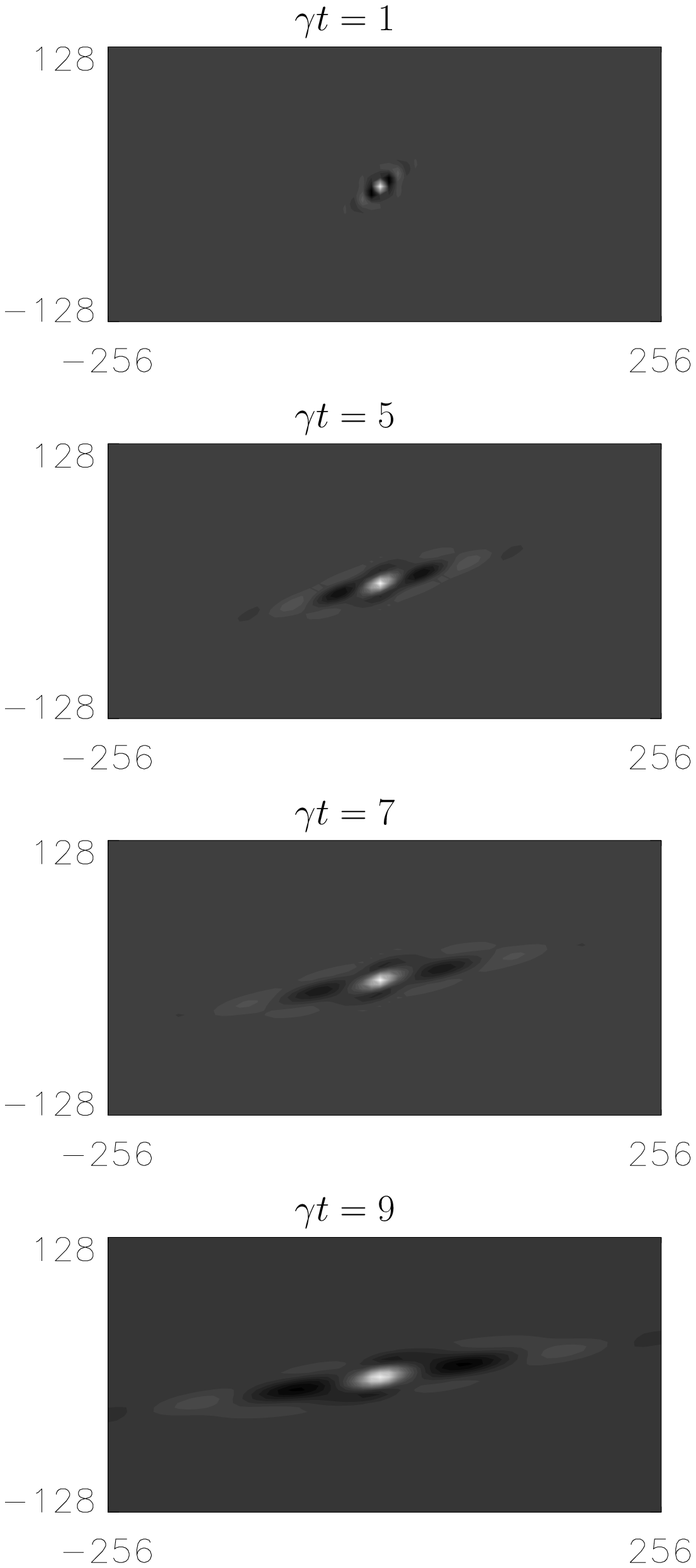,width=14cm,angle=0}}
\caption{The real-space correlation function $G(\vec r,t)$ is shown
(only a part of the whole lattice) 
at four different values of the strain, $\gamma t=1,5,7,9$. 
$x$ is on the horizontal axis.}
\label{figgr}
\end{figure}

The anisotropic character of the evolution, with the ordered regions
aligning along the flow direction, is reflected in the behavior
of the correlation function $G$, shown in Fig.~(\ref{figgr}).
$G$ has a peak at $\vec r=0$ and is continuously stretched along
the $x$-direction, assumining increasingly eccentric elliptical patterns. 

\begin{figure}
\centerline{\psfig{figure=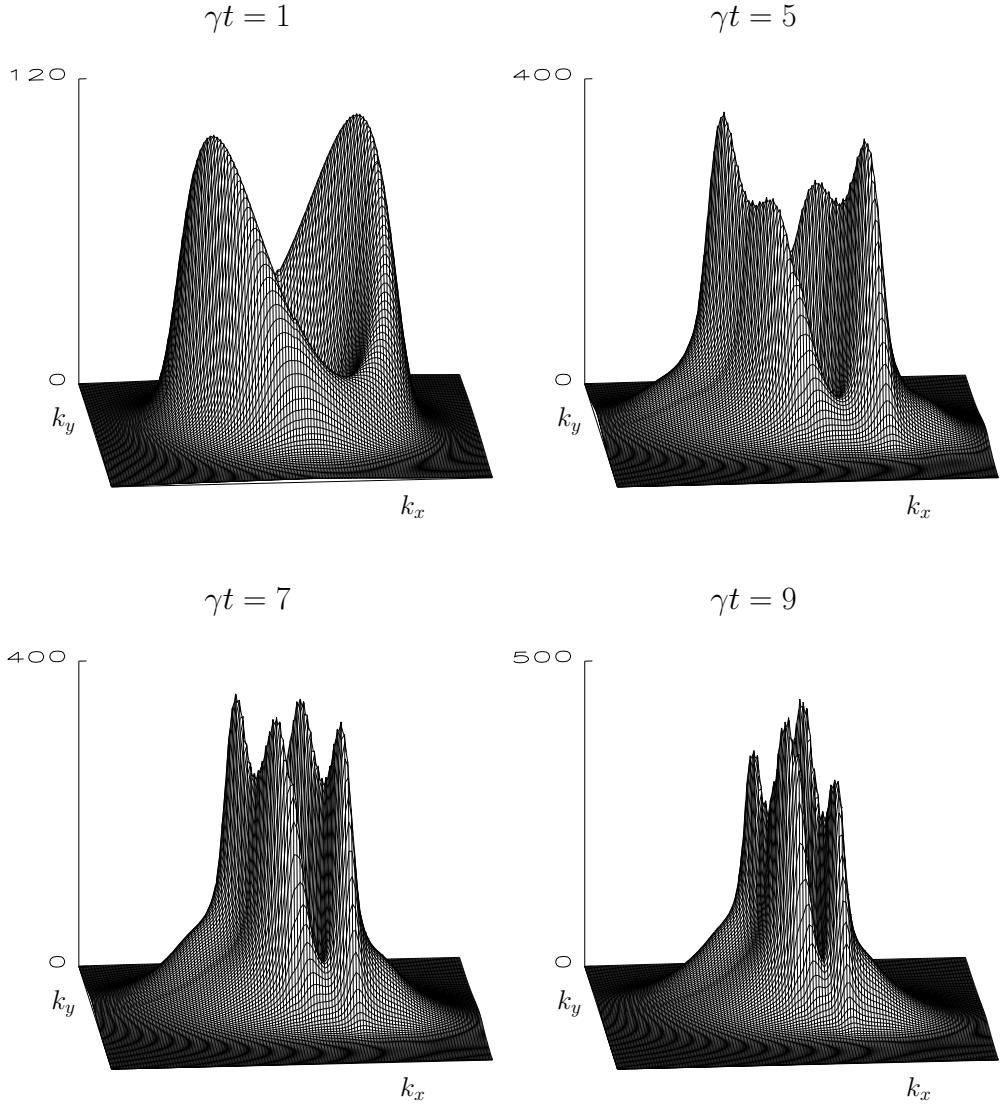,width=15cm,angle=0}}
\caption{The structure factor $C(\vec k,t)$ is shown
at four different values of the strain, $\gamma t=1,5,7,9$. 
$k_x$, on the horizontal
axis, increases from left to right; $ky$ increases toward the
back of the figure.}
\label{figck}
\end{figure}

Snapshots of the structure factor at the same times are shown in the $k_x,k_y$
plane  in Fig.~(\ref{figck}). 
Initially, not shown here,
$C$ develops the form of a circular volcano, as in the case without
shear. A later observation, at $\gamma t=1$,
shows a deep in the volcano profile along the direction $k_y= -k_x $.
This deep becomes progressively more pronounced until $C$ appears
separated into two symmetric foils.
At times of order $\gamma t=5$ in each foil a couple of maxima
are built up. We denote by the letter A the peak that in the figure
at time $\gamma t=5$ is higher. With respect to the other maximum,
denoted by B, A is characterized by having $\vert k_x \vert >>
\vert k_y \vert$, whereas in B $\vert k_x\vert \simeq \vert k_y \vert $.
The relative height of these maxima changes in time. This is 
observable in Fig.~(\ref{figck}) at time $\gamma t=9$: now 
B is prevailing. 

Up to this time the qualitative evolution of the structure factor 
is comparable to what reported for $N=\infty $ \cite{noiinf}
and for the numerical solution \cite{noinum} of the full model equation
with $N=1$. Computer limitations
prevent the observation of the system on longer times. 
Only in the $N=\infty $ case the dynamics can be
followed by numerical integration of the equations on larger 
timescales and it can
be shown that the repeated prevalence of either A or B maxima
continues cyclically in time. It turns also out that the 
recurrent dominance of A and B is periodic
in the logarithm of the strain $\gamma t$.
A physical interpretation of this pattern in terms of
stretching and break-up of domains is proposed in \cite{noinum}.

\begin{figure}
\centerline{\psfig{figure=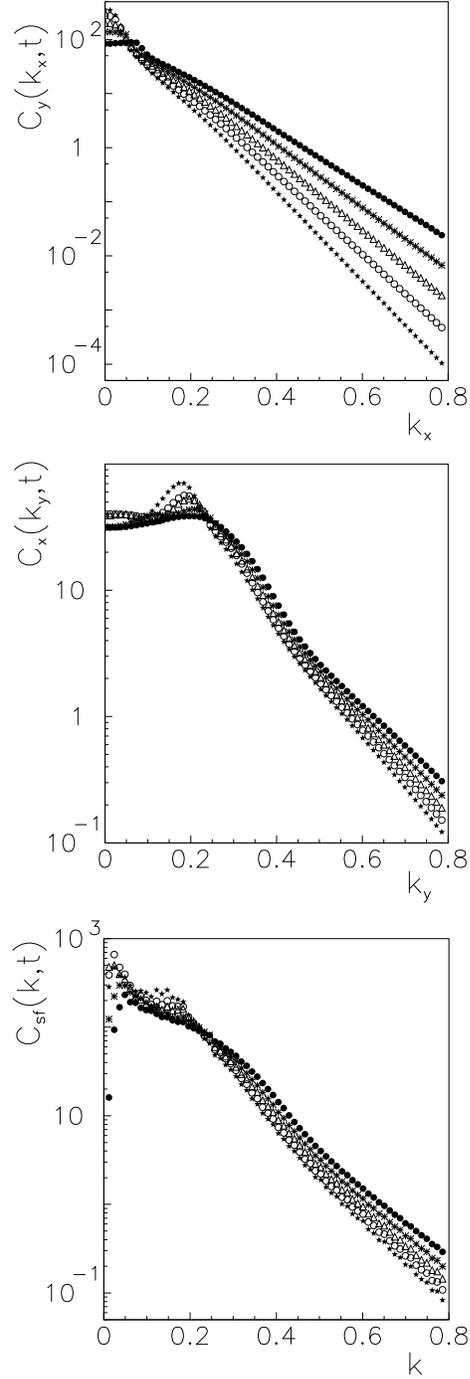,width=15cm,angle=0}}
\caption{The structure factors averaged along the $k_x$ and $k_y$ directions
and spherically are shown for $\gamma t=10$ 
($\bullet $), 15 ($\ast $), 20 ($\triangle $), 25 ($\circ $) and 30 ($\star $).} 
\label{figmedie}
\end{figure}

The spherical average
\be
C_{sf}(k,t)={\cal N}^{-1}\int d\theta C(\vec k,t),
\label{sf}
\ee
where $\tan \theta=k_y/k_x$ and ${\cal N}$ is the number of
lattice points contained in a circular shell of width $\pi /256$ centered 
around $k$, and the averages along $k_x$ or $k_y$
\be
C_x(k_y,t)=L^{-1}\int dk_x C(\vec k,t)
\label{cx}
\ee
and
\be
C_y(k_x,t)=L^{-1}\int dk_y C(\vec k,t)
\label{cy}
\ee  
are plotted in Fig.~(\ref{figmedie}). 
This figure shows that in the large-$\vec k$ tail the
structure factor decays exponentially, as already observed without
imposed flow~\cite{zancast}, at variance with the faster
decay observed for $N=\infty$~\cite{brayninf}.
As already mentioned, the absence of power-law tails
is related to the absence of stable localized topological defects
in the Bray-Humayun approximation. 

Next we consider the growth laws.
For $N=\infty $, in the large time region \cite{brayninf,noiinf}
$L_x$ and $L_y$ keep increasing with (logarithmically corrected) 
power laws $L_x\sim \gamma (t^5/\ln t)^{1/4}$, 
$L_y\sim (t/\ln t)^{1/4}$ modulated by a log-periodic
oscillation. The short time part of this pattern is observed
in Fig.~(\ref{figradii}b). The same growth exponents are expected
for the BH model (but without logarithmic corrections); this is
because the model amounts to an $1/N$ perturbative expansion and,
therefore, it is accurate for vectorial systems whose growth
exponents belong to the same universality class of $N=\infty $.
Actually the value of the exponents can be easily determined
by means of a scaling analysis. If standard scaling holds
the structure factor can be cast as
\be
C(\vec k,t)=L_x(t)L_y(t)f(X,Y)
\label{scalc}
\ee
where $X=k_xL_x(t)$ and $Y=k_yL_y(t)$ and $f$ is a scaling
function. From~(\ref{scalc}) one also has
\be
D(\vec k,t) =b L_x(t)L_y(t)h(X,Y)   
\label{scald}
\ee
where $b$ is a constant and $h$ another scaling function.
Inserting expressions~(\ref{scalc},\ref{scald}) into the equation
of motion of $C$~(\ref{ck}) and assuming that asymptotically
$L_x$ prevails over $L_y$ one obtains
\begin{eqnarray}
2Y^4f(X,Y) &=&
-2R(t)L_y(t)^2\left [ f(X,Y)+\frac{b}{N}h(X,Y)\right ]Y^2
+\frac{L_y(t)^5}{L_x(t)}\gamma X\frac{\partial f(X,Y)}{\partial Y} \nonumber \\
&-&
\frac{1}{4}\frac{dL_y(t)^4}{dt}
\left \{
1+\frac{\frac{d \left [\ln L_x(t) \right ]}{dt}}
{\frac{d\left [\ln L_y(t)\right ]}{dt}} \right \}
\label{inferscal}
\end{eqnarray}
The l.h.s. of Eq.~(\ref{inferscal}) does not depend explicitly
either on time and $\gamma $. 
Then in order to fulfill this equation for each $X,Y$
both $t$ and $\gamma $ must drop out from the r.h.s. Then one has
\be
\left \{ \begin{array}{ll}
                  L_x(t)\sim \gamma t^{\frac{5}{4}} \\
                  L_y(t)\sim t^{\frac{1}{4}} \\
                  R(t)\sim t^{-\frac{1}{2}}	
	  \end{array}
              \right .
\label{exponents}
\ee  
As already anticipated the exponents are those pertaining to systems
with continuous symmetry. Furthermore one can observe that the shear
rate enters linearly in $L_x$ but the growth in the shear direction $y$
is unaffected by $\gamma $. For completeness we recall that in the
full scalar model one expects different exponents, namely
$L_x\sim \gamma t^{4/3}$, $L_y\sim t^{1/3}$ from scaling \cite{brayninf}
or renormalization group \cite{noinum}
arguments, but these are not yet clearly observed in simulations. Moreover
the very existence of an asymptotic scaling dynamics is debated in this case.  
Fig.~(\ref{figradii}a) shows that initially
both $L_x$ and $L_y$ start growing with an exponent which is consistent
with the value $1/4$ as in absence of shear. In this time regime 
the fluid essentially does not feel the presence of the flow, as
already discussed for the structure factor.
Then, starting from $\gamma t \simeq 0.5$, $L_y$ keeps growing
unexpectedly larger than $L_x$ untill at $\gamma t\simeq 2$ it
decreases. Up to this point the BH model roughly resembles the
$N=\infty $ case, despite the fact that the overshoot of $L_y$
is much more pronounced in the latter case. Fig.~(\ref{figradii}c),
instead, shows that the full scalar model behaves quite differently,
with a pronounced decrease of $L_y$ in the range of times where 
in the other models a fast growth was observed. For longer times
the asymptotic stage is entered. The numerical integration of the BH model 
presented insofar shows that Eq.~(\ref{exponents})
is obeyed but, at variance with the other two cases, the oscillatory
pattern is absent or, at least, strongly depressed. 

Now we turn to the discussion of Fig.~(\ref{figexc}) where
the behavior of the excess viscosity is reported.
In all the cases considered $\Delta \eta $ initially grows to a maximum, 
then falls negative and raises to a second maximum. This pattern 
has been observed also in experiments \cite{osc} and is referred
to as {\it double overshoot}.
After the second maximum the asymptotic stage is entered.
On the basis of scaling arguments one expects that $\Delta \eta $
scales as $\gamma ^{-1} $ times the inverse of the typical volume 
$L_xL_y$ of the domains of the two species.
Then $\Delta \eta \sim \gamma ^{-2}t^{-3/2}$ for vector order parameter
and $\Delta \eta \sim \gamma ^{-2}t^{-5/3}$ for the scalar model.
The former exponent is observed in Fig.~(\ref{figexc}a) for
the BH model, although in the very last part of the simulation.
For $N=\infty $ the same exponent (with logarithmic corrections)
has been exactly computed \cite{brayninf}
and observed numerically 
on larger timescales, but modulated by log-time 
periodic oscillations  \cite{noiinf}.      
In the scalar case a definite exponent cannot be extracted from simulations,
but the oscillations are observed.  

\section{conclusions} \label{conclusions}

In this Article we have considered the behavior of a binary
fluid quenched below the mixing temperature in the presence
of an applied shear flow.
The description of this system has been carried out
in the context of the continuum convection-diffusion equation
based on the Ginzburg-Landau equilibrium free energy.
The nature of our approach, which neglects hydrodynamic
effects, limits the domain of applicability of the model to 
the diffusive regime which takes place at short times.
We have studied this model in the 
approximation scheme introduced by Bray and Humayun.
This amounts to an expansion in $1/N$ on top of the auxiliary
field theory approximation originally introduced by Mazenko.
A comparison is also presented with two different approaches:
the $N=\infty $ limit and the full scalar model, which is
the most appropriate for fluids. 
The first has the great advantage to be fully soluble analytically,
al least in the large time domain, providing an unambiguous
description. Furthermore it can be easily
studied numerically from the instant of the quench onwards.
However, besides the fact that the case $N=\infty $ is far from physical,
it has also proved to violate the qualitative symmetry of standard scaling
in favor of multiscaling, rendering this model insecure 
as an approximation to real systems.  
On the other hand the full scalar model is well suited for describing
fluids but it cannot be studied analytically. Numerical simulations,
although important, have not yet provided clear evidences
of the basic phenomenology, due to technical limitations.
The BH model is ideally located in between these two: It shares with
the large-$N$ limit the property of being a closed equation for
an observable, the correlation function, that is already an
ensemble average. Then, even if one resorts in the end to a numerical
solution, one still has the great advantage 
to avoid the average over a large number of realizations.
Secondly, due to its perturbative character, it is believed to represent
a step forward with respect to the case $N=\infty $.
This statement has been proven to be correct at least without shear.

\begin{figure}
\centerline{\psfig{figure=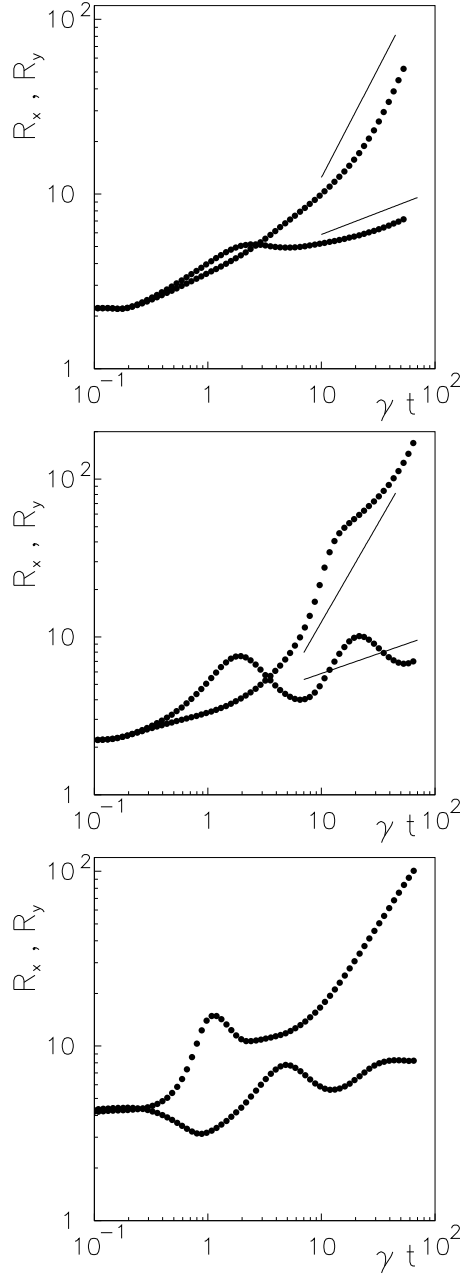,width=14cm,angle=0}}
\caption{The evolution of the characteristic lengths $L_x(t),L_y(t)$
is shown for a) the BH model, b) the $N=\infty $ limit, 
c) the full scalar model (numerical simulation). In a) and b)
the straight lines represent the expected power law behaviors
$L_x(t)\sim t^{5/4}$, $L_y(t)\sim t^{1/4}$.}
\label{figradii}
\end{figure}

\begin{figure}
\centerline{\psfig{figure=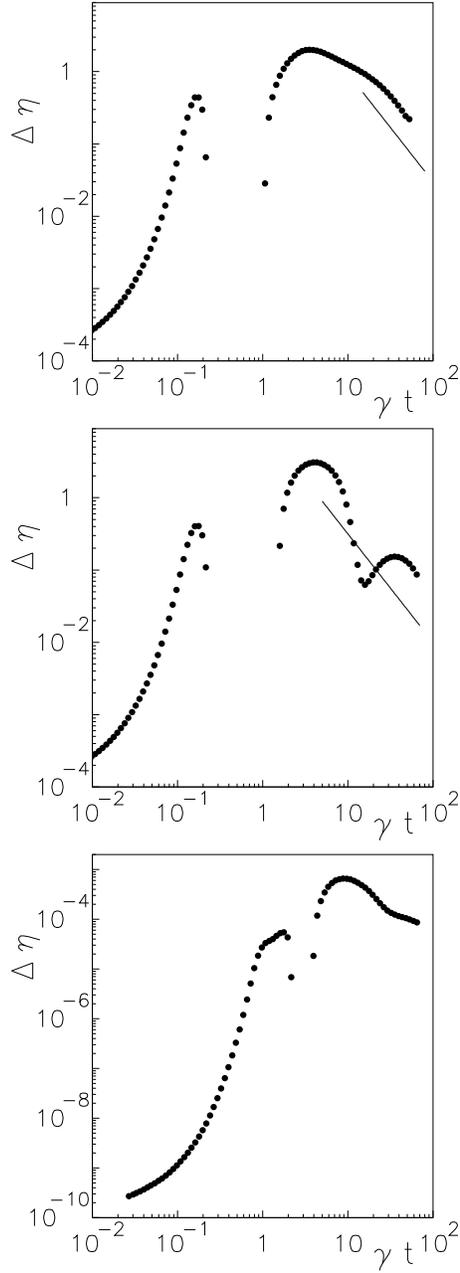,width=14cm,angle=0}}
\caption{The evolution of the excess viscosity $\Delta \eta (t)$
is shown for a) the BH model, b) the $N=\infty $ limit, 
c) the full scalar model (numerical simulation). 
The straight lines represent the power law behavior
$\Delta \eta(t)\sim t^{-3/2}$.}
\label{figexc}
\end{figure}

Given the essence of the BH scheme, 
one does not expect to obtain a fully reliable
approximation to real systems with
$N=1$. The spirit of this work is more to establish if 
and to which extent its
phenomenology compares better to the numerical simulations
of the physical model than to the case $N=\infty $. From 
the analysis of the behavior of the characteristic lengths 
and of the excess viscosity we 
can conclude that the BH model behaves more like the $N=\infty $ case with
damped oscillations than like the full scalar model, 
despite that, with the choice $N=1$, the corrections 
to the case $N=\infty $ contained in Eq.~(\ref{bh}) are effective
almost from the beginning, as it is clear from 
Figs.~(\ref{figradii},\ref{figexc}).
This is true not only for the value of the dynamic exponents,
but also for the qualitative behavior of the typical lengths.  

On the other hand, regarding the oscillating character of most
observables, the BH model behaves rather differently from both
$N=\infty $ and $N=1$ in that strong oscillations are observed
in the latter two while they  are practically absent in the former.
Under this respect the BH model is atypical and cannot be regarded
as a bridge between $N=\infty $ and $N=1$. It is possible that
the BH scheme resembles more the behavior of vectorial system with a
finite $N$ under shear, but this is a totally unexplored field.

Regarding the important issue of the existence of an asymptotic scaling
regime characterized by ever growing lengths
$L_x,L_y$, in the time range accessed in this
work we do not see any saturation effect and both $L_x$ and $L_y$ keep
firmly growing with the expected power laws. In the BH scheme, then, 
the presence of an asymptotic scaling regime is confirmed while
a stationary state with domains of a finite thickness is not observed.

{\bf Acknowledgments}

We warmly acknowledge Marco Zannetti for suggesting
this study.
F.C. and G.G. acknowledge partial
support by the European TMR Network-Fractals Contract
No. FMRXCT980183 and by INFM PRA-HOP 1999.

\newpage

\newpage

\newpage

\end{document}